# Vortices and Superfluidity in a Strongly Interacting Fermi Gas

M.W. Zwierlein, J.R. Abo-Shaeer*, A. Schirotzek, C.H. Schunck, and W. Ketterle

*Department of Physics, MIT-Harvard Center for Ultracold Atoms, and Research Laboratory of Electronics, MIT, Cambridge, MA 02139*

**Quantum-degenerate Fermi gases provide a remarkable opportunity to study strongly interacting fermions. In contrast to other Fermi systems, such as superconductors, neutron stars or the quark-gluon plasma, these gases have low densities and their interactions can be precisely controlled over an enormous range. Here we report observations of vortices in such a gas that provide definitive evidence for superfluidity. By varying the pairing strength between two fermions near a Feshbach resonance, one can explore the crossover from a Bose-Einstein condensate (BEC) of molecules to a Bardeen-Cooper-Schrieffer (BCS) superfluid of loosely bound pairs whose size is comparable to, or even larger than, the interparticle spacing. The crossover realizes a novel form of high-$T_C$ superfluidity and it may provide new insight for high-$T_C$ superconductors. Previous experiments with Fermi gases have revealed condensation of fermion pairs. While these and other studies were consistent with predictions assuming superfluidity, the smoking gun for superfluid behavior has been elusive. Our observation of vortex lattices directly displays superfluid flow in a strongly interacting, rotating Fermi gas.**

The first observation of Bose-Einstein condensates of molecules consisting of loosely bound fermionic atoms[1-3] initiated a series of explorations[4-12] of the BEC-BCS crossover[13-15]. When an external magnetic field is varied across a Feshbach resonance, these molecules transform adiabatically into the Cooper pairs of a BCS superfluid. All physical properties are expected to vary smoothly throughout this crossover. For example, the size of fermion pair condensates smoothly increases from the BEC- to the BCS-side of the resonance, while the strength of the bond between two paired atoms smoothly decreases. The similar size and shape of normal and condensed clouds makes it difficult to detect condensation on the BCS-side. However, using a rapid magnetic field sweep back to the BEC-side, introduced by the Boulder group[4], pair condensation was observed [4,5,12]. Although Bose-Einstein condensation and superfluidity are intimately connected, they do not necessarily occur together. In lower dimensions, superfluidity occurs in the absence of BEC[16]. An ideal Bose gas or disordered 3D system has a condensate at zero temperature, but shows no superfluidity[16]. Phase-fluctuations, which are likely present in short-lived condensates of fermionic pairs[1], can suppress superfluid behavior.

Several ground-breaking studies in Fermi gases of hydrodynamic expansion[7,17], collective excitations[8,9], thermodynamic properties[11] and the binding energy of pairs[10] were suggestive of superfluid behavior or were consistent with theoretical calculations predicting superfluidity, but did not provide unambiguous evidence. In the meantime, several theoretical papers[18-22] emphasized that the rotational properties of a gas of fermion pairs could directly reveal superfluidity in such systems.

Quantized vortices in a rotating gas provide conclusive evidence for superfluidity because they are a direct consequence of the existence of a macroscopic wavefunction that describes the superfluid. The velocity field of the superfluid is proportional to the gradient of the wavefunction's phase. In such a case, flow must be irrotational and angular momentum can enter the system only in the form of discrete line defects (vortices). In contrast, for a normal gas the lowest state of rotation corresponds to rigid body rotation. Metastable vortex patterns have been observed in classical inviscid fluids[23]. However, the final number and charge of the vortices depends chaotically on the initial conditions in contrast to the regular vortex lattices which we have reproducibly observed. Furthermore, vortex patterns in classical fluids are only stable at extremely low viscosity (i.e. for Reynolds numbers $> 10^5$). For a Boltzmann gas rotating close to the trap frequency the Reynolds number is approximately the cloud-size divided by the mean-free path, which does not exceed $10^3$ in our case. Pauli blocking can only decrease the Reynolds number further.

**Experimental procedure**

To create a strongly interacting Fermi gas, spin-polarized fermionic $^6$Li atoms were sympathetically cooled to degeneracy by $^{23}$Na atoms in a magnetic trap[24]. The Fermi cloud was then loaded into an optical dipole trap, and an 875 G external magnetic field was applied. Here a 50% - 50% spin-mixture of the two lowest hyperfine states of $^6$Li was prepared. Between these two states, labelled $|1\rangle$ and $|2\rangle$, there is a 300 G wide Feshbach resonance located at 834 G [25,26]. Evaporative cooling (achieved by reducing the laser power) accompanied by a magnetic field ramp to 766 G on the BEC-side of the resonance, typically produced a Bose-Einstein condensate of $3\times10^6$ molecules[3].

Previous experiments studying the rotation of *atomic* Bose-Einstein condensates employed magnetic traps operating at low bias fields[27-31]. Because the Feshbach resonance in our system occurs between two high-field seeking states that cannot be trapped magnetically, an optical dipole trap operating at high magnetic bias fields was necessary. Our setup employed a trapping beam with a $1/e^2$ radius of 123 μm (wavelength 1064 nm), radially confining the gas with a trap frequency of 59 Hz at a power of 145 mW. Axial confinement with trap frequency $v_z = 23$ Hz was provided by an applied magnetic field curvature that decreased the radial trap frequency to $v_r = 57$ Hz. The aspect ratio of the trap was 2.5. In this trap, at a field of 766 G, condensates of $1\times10^6$ molecules (the typical number in our experiment after rotating the cloud) have Thomas-Fermi radii of about 45 μm radially and 110 μm axially, a peak molecular density of $2.6\times10^{12}$ cm$^{-3}$, a chemical potential of about 200 nK, and a characteristic microscopic length scale of $1/k_F \approx 0.3$ μm. Here, the Fermi wavevector $k_F$ is defined by the Fermi energy of a non-interacting two-state mixture of atoms with total atom number $N$ in a harmonic trap of (geometric) mean frequency $\bar{\omega}$, $E_F = \hbar\bar{\omega}(3N)^{1/3} \equiv \hbar^2 k_F^2/2m$. Throughout this paper we will estimate the interaction parameter $1/k_F a$ using the average number of fermion pairs $N/2 = 1\times10^6$. Here, $a$ is the scattering length between atoms in states $|1\rangle$ and $|2\rangle$. At a field of 766 G, $1/k_F a = 1.3$. Because this gas is strongly interacting, it is difficult to extract a temperature from the spatial profile. For weaker interactions (at 735 G) the condensate fraction was in excess of 80%, which would isentropically connect to an ideal Fermi gas[32] at $T/T_F = 0.07$. The BEC-BCS crossover ($1/k_F |a| < 1$) occurs in the region between 780 G and 925 G.

The trapped cloud was rotated about its long axis using a blue-detuned laser beam (532 nm)[28,29,33]. A two-axis acousto-optic deflector generated a two-beam pattern (beam separation $d = 60$ μm, Gaussian beam waist $w = 16$ μm) that was rotated symmetrically around the cloud at a variable angular frequency Ω. The two beams with 0.4 mW power each produced a repulsive potential of 125 nK for the $^6$Li cloud, creating a strongly anisotropic potential. This method was first tested using a weakly interacting, *atomic* BEC of $^{23}$Na in the $|F = 2, m_F = 2\rangle$ state in an optical trap with $v_r = 60\,\text{Hz}$, $v_x = 23\,\text{Hz}$. Fully equilibrated lattices of up to 80 vortices were observed. The vortex number decayed with a $1/e$ lifetime of $4.2 \pm 0.2\,\text{s}$, while the atom number decayed, due to three-body losses and evaporation, with a lifetime of $8.8 \pm 0.4\,\text{s}$. The roundness of the optical trap and its alignment with both the optical stirrer and the axes of the magnetic potential were critical. Any deviation from cylindrical symmetry due to misalignment, optical aberrations, or gravity rapidly damped the rotation. The generation of vortices in sodium was comparatively forgiving and had to be optimized before vortices in $^6$Li$_2$ could be observed.

**Observation of vortex lattices**

In experiments with $^6$Li close to the Feshbach resonance, the interaction strength between atoms in states $|1\rangle$ and $|2\rangle$ can be freely tuned via the magnetic field. Thus, it is possible to choose different magnetic fields to optimize the three steps involved in the creation of a vortex lattice: Stirring of the cloud (for 800 ms at a typical stirring frequency of 45 Hz), the subsequent equilibration (typically 500 ms) and time-of-flight expansion for imaging. To stay close to the analog case of an atomic condensate, our search for vortices started on the BEC-side of the resonance, at a fixed magnetic field of 766 G. To image the cloud, the trapping beam was switched off and the cloud expanded in the residual magnetic potential. After 12 ms time-of-flight, a molecular absorption image along the axial direction was taken with light resonantly exciting atoms in state $|2\rangle$ [3] (see Fig. 1a). While the contrast of the vortex cores was low, a regular lattice pattern containing about 25 vortices was visible. This establishes superfluidity for molecular condensates.

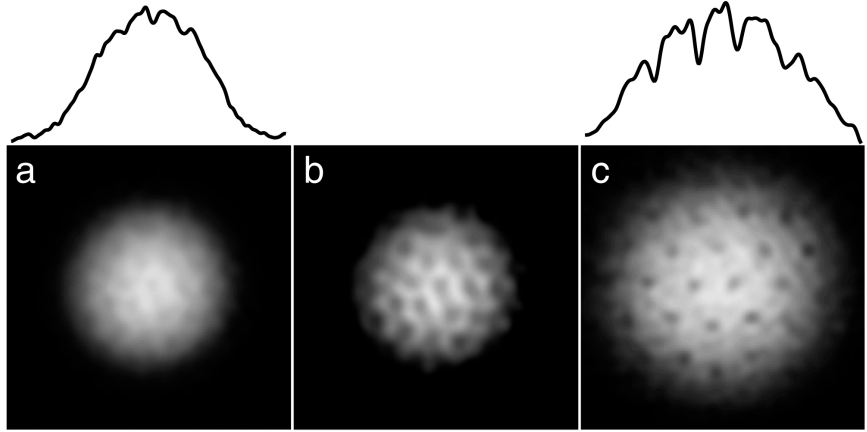

Fig. 1: Observation of a vortex lattice in a molecular condensate. (a) Fixed field. Stirring for 800 ms, followed by 400 ms of equilibration, and imaging after 12 ms time-of-flight all took place at 766 G. The vortex core depletion of the integrated density profile is barely 10%, as indicated by the 5-$\mu$m-wide cut on top. (b) Fourier-filter applied to (a) to accentuate the vortex contrast. Spatial frequencies with an absolute value of about the inverse vortex core size were enhanced by a factor of four. (c) Varying field. The vortex lattice was created at 766 G and imaged at 735 G following the procedure outlined in the text. The vortex core depletion is now about 35%. The field of view is $780\ \mu m \times 780\ \mu m$.

Subsequently, it was found that the contrast of the vortex cores could be enhanced by the following steps: The magnetic field curvature was reduced by a factor of five during the first ms of time-of-flight. After 2 ms expansion at the initial field, the magnetic field was ramped down over 2 ms to 735 G. The cloud was imaged after additional 9 ms of expansion at this field. Attempts to image at even lower fields did not enhance the contrast. Further improvements were achieved by forcing the cloud to expand faster by increasing the power of the optical trap by a factor of 4.5 during the last 2 ms of trapping (see Fig. 1c). We suspect that due to the residual magnetic field curvature, a faster expansion was superior to longer time-of-flight.

Using this procedure we observed vortices that were created not only on the BEC- but also the BCS-side of the Feshbach resonance, at magnetic fields between 740 G and 863 G (Fig. 2). On the BCS-side, isolated fermion pairs are unstable. As the cloud expands from the trap and the density decreases, the pairs will become more fragile and dissociate at a certain point in time-of-flight. Information on the center-of-mass wavefunction of the pairs, and hence the vortex contrast, will be gradually lost. The field ramp to the BEC-side during expansion protects the pairs by transforming them into stable molecules. At 853 G, this ramp could be delayed by 6 ms into time-of-flight before the vortex contrast was lost. It is not clear why this ramp was found to be necessary already at 812 G, on the BEC-side of the resonance.

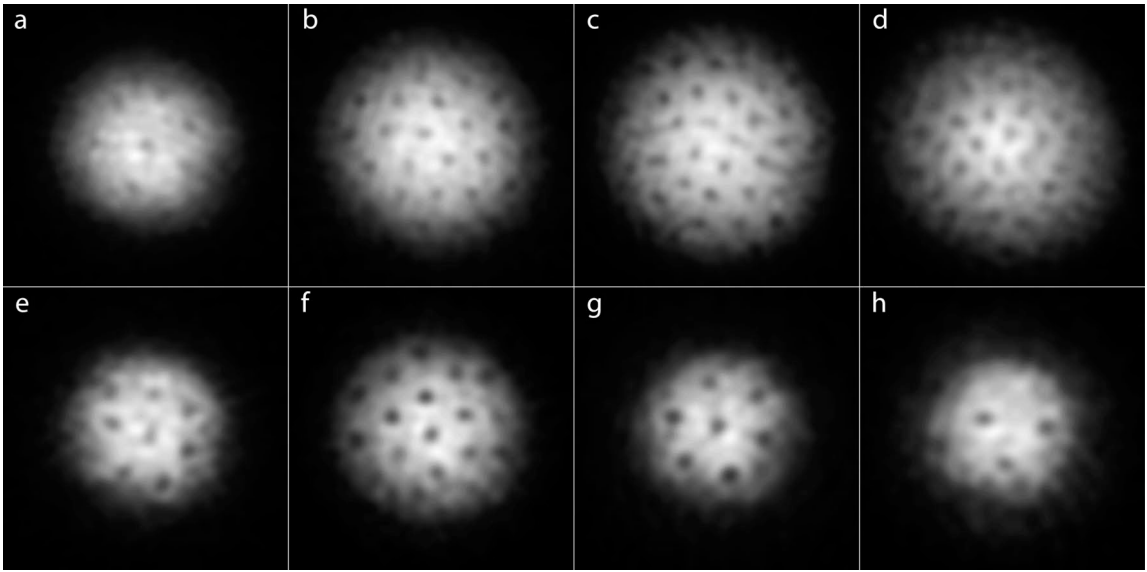

Fig. 2: Vortices in a strongly interacting gas of fermionic atoms on the BEC- and the BCS-side of the Feshbach resonance. At the given field, the cloud of lithium atoms was stirred for 300 ms (a) to 500 ms (b-h) followed by an equilibration time of 500 ms. After 2 ms of ballistic expansion, the magnetic field was ramped to 735 G for imaging (see text for details). The magnetic fields were (a) 740 G, (b) 766 G, (c) 792 G, (d) 812 G, (e) 833 G, (f) 843 G, (g) 853 G and (h) 863 G. The field of view of each image is $880\ \mu m \times 880\ \mu m$.

It is impossible for vortex lattices to form during ballistic expansion at the imaging field (735 G). We show below that the formation of vortex lattices even at the high density of the trapped cloud is several hundred milliseconds. Furthermore, even if there was some unpredicted fast formation mechanism for vortices, they could not form a regular array with long-range order in a cloud which expands at the speed of sound of the trapped gas. The observation of vortex lattices on the BCS-side of the Feshbach resonance above 834 G demonstrates superfluidity of fermions at magnetic fields where they cannot form stable, isolated molecules.

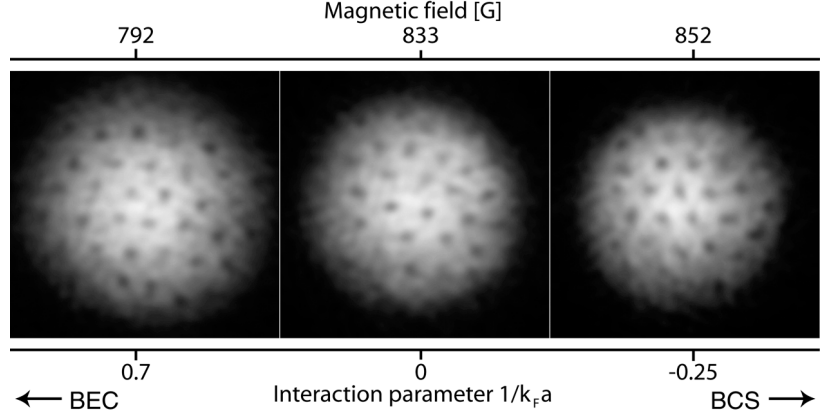

Fig. 3: Optimized vortex lattices in the BEC-BCS crossover. After a vortex lattice was created at 812 G, the field was ramped in 100 ms to 792 G (BEC-side), 833 G (resonance), and 853 G (BCS-side), where the cloud was held for 50 ms. After 2 ms of ballistic expansion, the magnetic field was ramped to 735 G for imaging (see text for details). The field of view of each image is 880 μm × 880 μm.

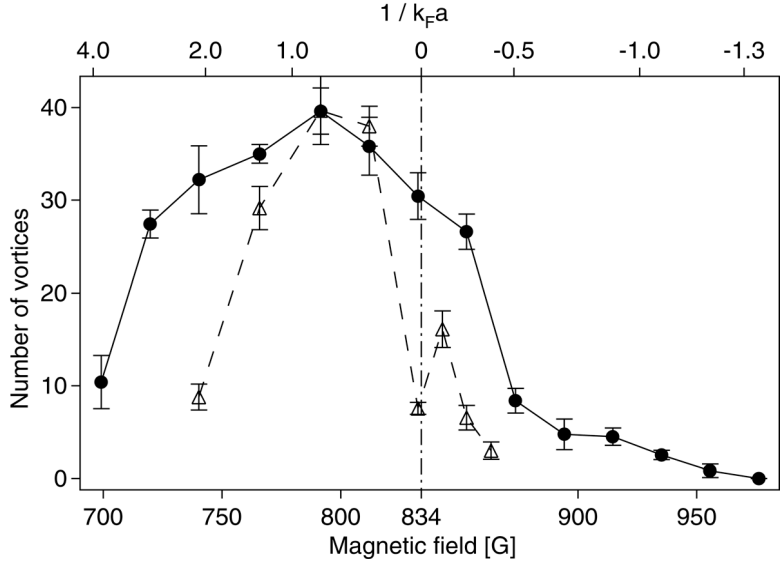

Fig. 4: Vortex number vs magnetic field in the BEC-BCS crossover. The open triangles show the number of vortices obtained after stirring and equilibration at the given, fixed magnetic field, as in Fig. 2. For the solid circles, a procedure similar to the one in Fig. 3 was used, where the vortex lattice was prepared at 812 G, and then the field was ramped to the test field. The position of the Feshbach resonance[26] is marked with the dash-dotted line. The data points and error bars give the average and standard deviation of several measurements.

The highest number of vortices (~ 40) was obtained by stirring at 766 G and then equilibrating close to resonance at 812 G ($1/k_F a = 0.35$). We suspect that the violent nature of the stirring produced more heating near the Feshbach resonance where the pairs are loosely bound. On the other hand, fields closer to resonance were favourable for equilibration due to suppression of vibrational relaxation. After preparing such a vortex lattice at 812 G, the magnetic field was ramped over 100 ms to a test field. After a hold time of 50 ms, the vortex lattice was imaged as discussed above (Fig. 3). Vortices were observed for test fields between 700 G ($1/k_F a = 3.8$) and 954 G ($1/k_F a = -1.2$) (Fig. 4).

The regularity of the lattice proves that all vortices have the same vorticity. From their number, the size of the cloud and the quantum of circulation $h/2m$ for each vortex, where $m$ is the mass of a lithium atom, we can estimate the rotational frequency of the lattice. For an optimized stirring procedure, we find that it is close to the stirring frequency. This excludes a quantum of circulation of $h/m$ or doubly charged vortices.

**Formation and lifetime of vortex lattices**

Before we found the detection scheme described above, we studied the formation and decay of the vortex lattices using a different procedure. The magnetic field was lowered to 735 G ($1/k_F a = 2.3$) already in the last 30 ms before expansion. Reduction of the magnetic field curvature by a factor of five took place during the last 5 ms before expansion avoiding undesired compression of the cloud in the axial direction. As before, the trap was compressed by increasing the trapping power by a factor of 4.5 during the last 2 ms before the switch-off. Imaging was done after 12 ms expansion at 735 G.

When a superfluid is rotated, the creation of quantized vortices is energetically favored only above a certain critical rotation frequency. In some cases a higher rotational frequency is necessary to actually nucleate the vortices[34]. This occurs through a dynamic instability of surface excitations[33,35,36]. Fig. 5 shows that vortices in the BEC region were created over a large range of stirring frequencies, as opposed to only near the quadrupole surface mode resonance[28,29]. Similar non-resonant behavior was observed in atomic sodium condensates with small stirring beams[36]. The strong dependence of the observed stirring efficiency curve on the magnetic field strength could be related to the increase of the speed of sound and hence the critical velocity for stronger interactions.

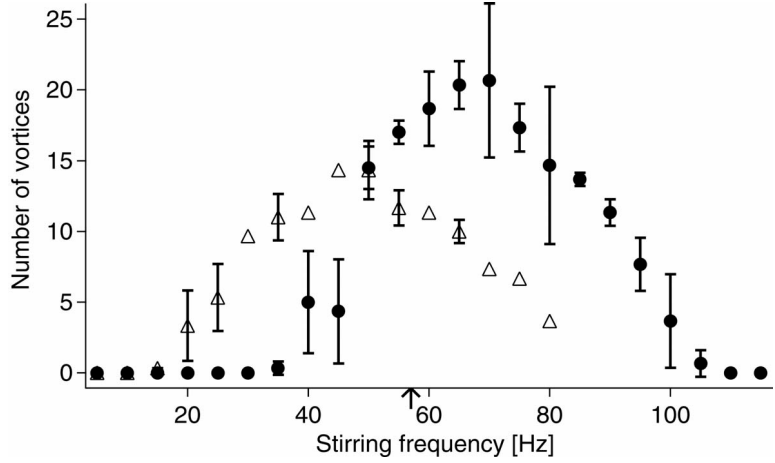

Fig. 5: Vortex number vs stirring frequency in the BEC region for different interaction strengths. Vortices were efficiently created over a broad range of stirring frequencies. The triangles and circles correspond to stirring (for 800 ms) and equilibration (for 500 ms) at 766 G and 812 G, respectively. The data points and error bars (if larger than the symbol) show the average and standard deviation of three measurements. The radial trap frequency (57 Hz) is indicated by the small arrow.

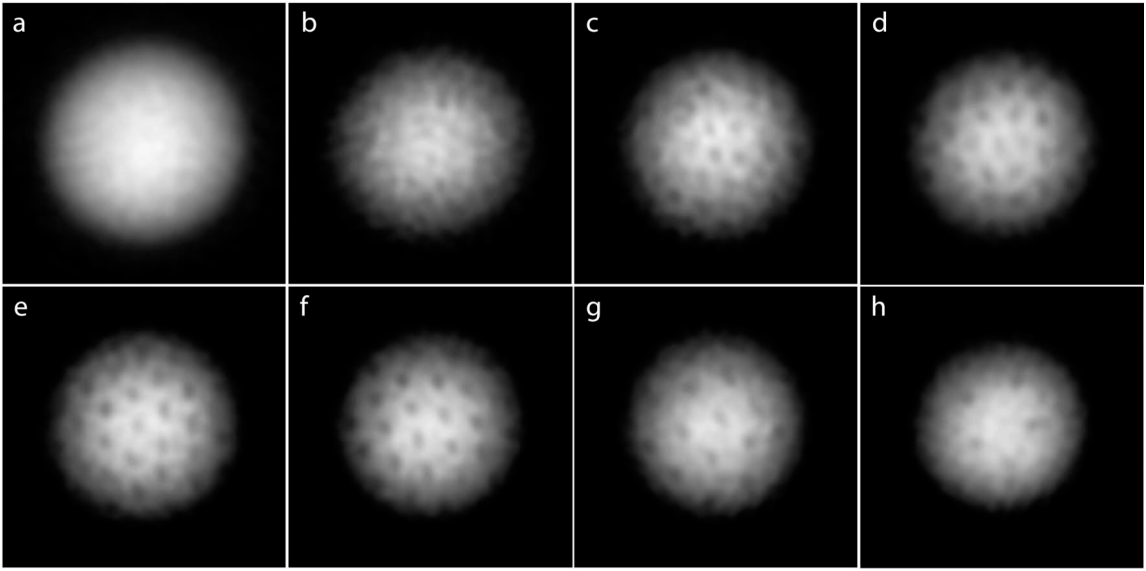

Fig. 6: Formation and decay of a vortex lattice in a fermion pair condensate on the BEC-side close to the Feshbach resonance. A molecular condensate, prepared at 766 G as shown in (a), was stirred for 800 ms. The field was then ramped to 812 G in 20 ms for equilibration. At this field, $1/k_F a = 0.35$, and the condensate was deep in the strongly interacting regime. To observe the vortex lattice, the field was ramped in 25 ms to 735 G ($1/k_F a = 2.3$), where the condensate was released from the trap and imaged after 12 ms time-of-flight. The equilibration times after the end of the stirring were (b) 40 ms, (c) 240 ms, (d) 390 ms, (e) 790 ms, (f) 1140 ms, (g) 1240 ms and (h) 2940 ms. Due to stirring, evaporation and vibrational relaxation, the number of fermion pairs decayed from $3 \times 10^6$ (a) to $1 \times 10^6$ (b-h). The field of view of each image is $830\ \mu m \times 830\ \mu m$.

Fig. 6 shows the formation of a vortex lattice close to the Feshbach resonance, on the BEC-side at 812 G ($1/k_F a = 0.35$). Immediately after stirring, the cloud was in a turbulent state (Fig. 6b). It took several hundred milliseconds for the vortices to fully crystallize into a lattice (Fig. 6c-e). The vortices arranged themselves in a hexagonal Abrikosov lattice to minimize their interaction energy[29,37]. Because the trap potential was not perfectly round (trap asymmetry $(v_x^2 - v_y^2)/(v_x^2 + v_y^2) \approx 0.03$), the vortex lattice slowly decayed on a timescale of several seconds (Fig. 6f-h). These observations are fully analogous to those already made in *atomic* Bose-Einstein condensates[29,35,38]. The formation time of several hundred milliseconds is in agreement with these experimental studies as well as with a recent theoretical study on the vortex lattice formation in a strongly interacting Fermi gas[39]. Note that this time scale was found to be independent of temperature[38] and seems to represent an intrinsic time scale of superfluid hydrodynamics. The long formation time excludes that ordered vortex lattices can be created during the 30 ms field ramp to 735 G, used for imaging. This holds even more strongly for the imaging method described first, where the magnetic field was switched only in time-of-flight. We are not aware of any possible formation mechanism that could create regular vortex lattices during ballistic expansion.

In principle, the presence of vortices can be used to map out the superfluid regime as a function of temperature and interaction strength, even in regions where fermion pairs are much larger than the interparticle spacing and can no longer be detected by transforming them into stable molecules[4,5,12]. As a first step, the lifetime of the vortex lattice was studied at different magnetic fields / interaction strengths. After preparing a fully crystallized vortex lattice containing about 30 vortices in the BEC-region at 812 G, the magnetic field was ramped to a chosen point in the crossover. After a variable hold time, the cloud was imaged at 735 G as described above, and the remaining number of vortices was counted. The results of this measurement are summarized in Fig. 7. The longest lifetime, 3.4 s, was obtained for magnetic fields slightly below resonance, near 810 G ($1/k_F a = 0.4$). Here, we observed 4 vortices even after 7 s. During this time the fluid at the vortex core (with characteristic size $1/k_F$) rotated more than 50,000 times, displaying truly superfluid behavior. As expected, the lifetime was reduced at low magnetic fields where the molecules heat up due to vibrational relaxation. In addition, and totally unexpected, a narrow dip in lifetime at approximately 831 G with a width of 8 G was observed. We speculate that this decrease in lifetime so close to resonance is caused by a coupling of the external motion of the loosely bound pairs to their internal motion, causing pair breaking or rotational excitation. Indeed, at 831 G, the binding energy of $\sim h \cdot 35$ Hz for isolated molecules is comparable to the initial rotational velocity of the lattice. The reduced lifetime on the BCS-side could be caused by an increasing fraction of thermal fermion pairs[5,38] due to the decreasing critical temperature for superfluidity. In a two-fluid model, angular momentum is stored as a vortex lattice in the superfluid component. Friction is provided by the normal component which increases as we move further away from resonance on the BCS-side. At 925 G ($1/k_F a = -1.0$) the vortex lifetime was reduced to only 10 ms, with a large error bar indicating strong fluctuations in the vortex number. Here we might be approaching the region where the superfluid to normal transition takes place.

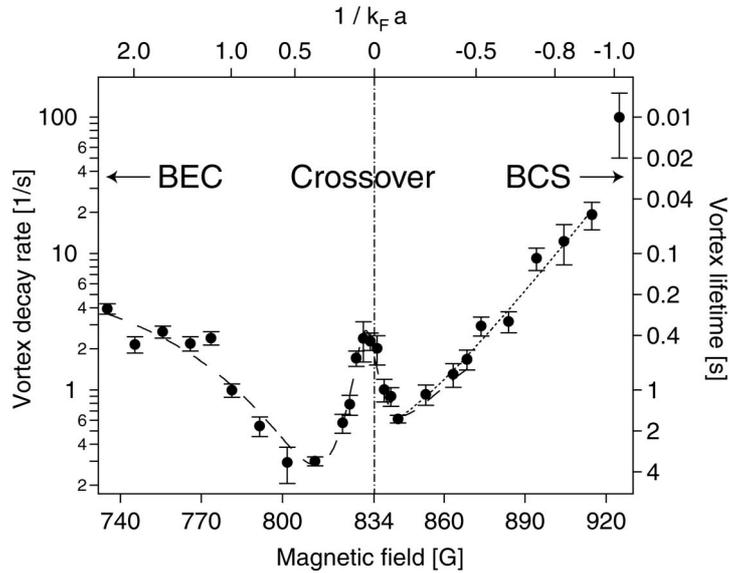

Fig. 7: Decay rate and lifetime of the vortex lattice vs magnetic field and interaction strength. The position of the Feshbach resonance[26] is marked with the dash-dotted line. The dashed and dotted lines are Gaussian and exponential fits to guide the eye, the dashed line including a lorentzian fit for the narrow feature near resonance.

## Conclusions

We have detected long-lived vortex lattices in a strongly-interacting Fermi gas over the entire BEC-BCS crossover region by imaging them after switching to lower magnetic fields during ballistic expansion. This provides the first direct signature of superfluidity in these systems. We expect that vortices in rotating Fermi gases will serve as an important starting point for future studies on superfluid dynamics.

We would like to thank Peter Zarth for experimental assistance and Claudiu Stan for important contributions in the early stages of the experiment. We would also like to acknowledge fruitful discussions with the participants of the OCTS conference in Ohio, and express our thanks to James Anglin, Zoran Hadzibabic, Dan Kleppner and Aaron Leanhardt for a critical reading of the manuscript. This work was supported by the NSF, ONR, ARO, and NASA.

Correspondence should be addressed to M.W.Z. (e-mail: zwierlei@mit.edu).

* present address: Lawrence Berkeley National Laboratory, One Cyclotron Road, MS 88R0192, Berkeley, CA 94720